\documentstyle[prd,aps,epsf]{revtex}
\begin{document}

\twocolumn[\hsize\textwidth\columnwidth\hsize\csname
@twocolumnfalse\endcsname

\title{Brane-World and Holography}

\author{Tetsuya Shiromizu$^{(a)}$, Takashi Torii$^{(a), (b)}$ 
and Daisuke Ida$^{(a)}$}

\address{$^{(a)}$ Research Centre for the Early Universe (RESCEU),
The University of Tokyo, Tokyo 113-0033, Japan}
\address{$^{(b)}$ Advanced Research Institute for Science and Engineering,
Waseda University,
Shinjuku-ku, Tokyo 169-8555, Japan}




\maketitle

\begin{abstract}
We consider the brane-world in the holographic point of view. 
Bearing the realistic models in mind, the bulk 
massless scalar field is introduced. First of all, we find the 
constraint on the coupling of the scalar fields with the 
matter(not holographic CFT) on the brane. We show that the 
traceless part of the energy-momentum tensor of 
holographic CFT is a part of the bulk Weyl tensor. The trace part which 
comes from the trace-anomaly is corresponding to 
the $\rho^2$-term appeared in the generalized FRW equation in the 
brane-world. 
\end{abstract}
\vskip2pc]

\vskip1cm

\section{Introduction}

The Randall-Sundrum brane-world is the attractive and simple model
which describes the stringy cosmology \cite{RS1,RS2}, 
where the 4-dimensional spacetime is regarded as a 4-dimensional 
Minkowski spacetime embedded in the 5-dimensional 
anti-de Sitter (AdS) spacetime. 
In this model,
it has been checked that the Newton gravity 
is recovered in the linearized theory \cite{RS1,RS2,Tama,Sasaki,Tess,Lisa},
and in special cases, it has 
recently been confirmed that the Einstein gravity 
is recovered up to the second order perturbation analysis \cite{Kudoh}. 
Due to the technical reason, however, 
the non-linear aspects of the brane-world are
still unclear (See Ref.~\cite{Non} for some attempts on this issue). 
It is remarkable that 
the homogeneous-isotropic universe is described by the domain wall 
motion in the 5-dimensional Schwarzschild-AdS spacetime \cite{cosmos,Misao}.

We shall consider in this paper that the gravitational equations on the brane 
from the holographic point of view. We will compare the gravitational 
equations, which are obtained via the purely geometrical reduction without any 
approximations, with the effective equations derived by using the 
AdS/CFT correspondence \cite{AdSCFT}. 
This direction was initiated 
by Witten, and then formulated by others \cite{Lisa,AdS1,Ida,John}. 
They have shown that 
the correction to the Newton gravity can be calculated by  CFT living 
on the brane(See Ref \cite{apply} for the cosmological applications). 
In connection with this topic, other aspects of the 
dilatonic brane-world have been actively investigated so far 
\cite{Lukas,Andrew,Wands,Richard,Cevetic,Seto}. 

In this paper we will consider what kind of brane-world 
is compatible to the AdS/CFT correspondence. There are two 
ways to obtain the effective equation on the brane. 
By comparing them, our purpose should be attained. 
The rest of this report is organised as follows. In Sec.~\ref{II}, 
following Refs. \cite{Tess,Wands}, we give a brief review of 
the gravitational equations on the brane. 
In Sec.~\ref{III}, we derive the effective equations on the brane 
via the AdS/CFT correspondence and compare them with those
described in Sec.~\ref{II}. To make the points clear, we  
consider a simple model there. The model, which 
can be described by the aspect of the AdS/CFT correspondence, has a 
certain constraint on the 
coupling of the scalar field to the matters on the brane. Finally, 
we give a summary in Sec.~\ref{IV}.

\section{The gravitational equation on the brane}\label{II}

In this section, we briefly review the gravitational equation on the 
brane derived in Refs. \cite{Tess,Wands}. We begin with the action
%
\begin{eqnarray}
S & = & S_{\rm bulk}+S_{\rm brane} \nonumber \\
& = & \Bigl\lbrace \int d^5 x {\sqrt{-g}} 
\Bigl[\frac{1}{2{\kappa_5}^2}{}^{(5)}R-\frac{1}{2}(\nabla\phi)^2
-V(\phi)   \Bigr]\nonumber \\
& & +\int d^4 x {\sqrt{-g}}\frac{1}{{\kappa_5}^2}K \Bigr\rbrace
\nonumber \\
& & +\int d^4 x {\sqrt{-q}}\Bigl[{\cal L}_{\rm m}-\lambda(\phi)\Bigr].
\label{begin} 
\end{eqnarray}
%
We here work on the Gaussian normal coordinate such that the metric takes the form
%
\begin{eqnarray}
ds^2=g_{MN}dx^M dx^N=dy^2+q_{\mu\nu}(y,x)dx^\mu dx^\nu.
\end{eqnarray}
%
The brane is assumed to locate at $y=0$ and 
$q_{\mu\nu}(0,x)=:q_{\mu\nu}(x)$ is the induced metric on the brane. 
{}From the Gauss equation and the Israel's junction condition \cite{Israel}, 
the gravitational equation on the brane is given by
%
\begin{eqnarray}
{}^{(4)}G_{\mu\nu} & = & \frac{2{\kappa_5}^2}{3}\hat T_{\mu\nu}(\phi)
\nonumber \\
& & +g_{\mu\nu}\Bigl[-\Lambda_4+\frac{{\kappa_5}^2}{16}
\Bigl(2\frac{d\lambda}{d\phi}+
\frac{\partial {\cal L}_m}{\partial \phi} 
\Bigr)  \frac{\partial {\cal L}_m}{\partial \phi } \Bigr] \nonumber \\
& & +8\pi G_4 (\phi)\tau_{\mu\nu}+{\kappa_5}^4\pi_{\mu\nu}-E_{\mu\nu},
\label{basic1}
\end{eqnarray}
%
where
%
\begin{eqnarray}
\hat T_{\mu\nu}(\phi)=D_\mu \phi D_\nu \phi 
-\frac{5}{8}q_{\mu\nu}(D\phi)^2,
\end{eqnarray}
%
%
\begin{eqnarray}
\Lambda_4=\frac{1}{2}{\kappa_5}^2\Bigl[V
+\frac{1}{6}{\kappa_5}^2 \lambda^2
-\frac{1}{8}\Bigl( \frac{d\lambda}{d\phi}\Bigr)^2 \Bigr],
\end{eqnarray}
%
%
\begin{eqnarray}
8\pi G_4 (\phi)=\frac{{\kappa_5}^4}{6}\lambda(\phi),
\end{eqnarray}
%
%
\begin{equation}
\pi_{\mu\nu}=-\frac14 \tau_{\mu\alpha}{\tau_{\nu}}^{\alpha}
+\frac{1}{12}\tau\tau_{\mu\nu}
+\frac18 q_{\mu\nu}\tau_{\alpha\beta}\tau^{\alpha\beta}
-\frac{1}{24}q_{\mu\nu}\tau^2,
\end{equation}
%
and
%
\begin{eqnarray}
E_{\mu\nu}={}^{(5)}C_{\mu M \nu N}n^M n^N.
\end{eqnarray}
%
$\tau_{\mu\nu}$ is the energy-momentum tensor of ${\cal L}_m$ 
on the brane. $D_\mu$ is the covariant derivative with respect to 
$q_{\mu\nu}$. 
Here we have taken into account a possibility that
the matter fields on the brane couple to the dilaton field $\phi$.
The second-rank trace-free symmetric tensor $E_{\mu\nu}$ 
is the so-called ``electric'' part of the 5-dimensional 
Weyl tensor, ${}^{(5)}C_{KLMN}$. Note that we did not use any approximation
to derive the gravitational equations. In this sense, Eq.~(\ref{basic1}) 
is exact under the assumption that the full 
action is given by Eq.~(\ref{begin}). 
Since $E_{\mu\nu}$ is the 5-dimensional quantity, however, 
the above equations are 
not a closed system in the 4-dimensional sense. To evaluate $E_{\mu\nu}$ 
on the brane, we must solve the equation for $E_{\mu\nu}$ in the bulk
\cite{Sasaki,Gen}. 

In a similar way, we obtain the equation for the scalar field on the brane:
%
\begin{equation}
D^2\phi-\frac{{\kappa_5}^2}{12}(4\lambda -\tau) 
\Bigl(\frac{d\lambda}{d\phi}+\frac{\partial {\cal L}_m}{\partial \phi} 
\Bigr)-\frac{dV}{d\phi}=-\partial_y^2 \phi|_{\rm brane}.
\end{equation}
%

The trace part of the gravitational equation on the brane will 
play an important role for the discussion in the 
next section:
%
\begin{eqnarray}
{}^{(4)}R & = & {\kappa_5}^2(D \phi)^2 \nonumber \\
& & +4 \Bigl[ \Lambda_4-\frac{{\kappa_5}^2}{16}
\Bigl(2\frac{d\lambda}{d\phi}+\frac{\partial {\cal L}_m}{\partial \phi} \Bigr)
\frac{\partial {\cal L}_m}{\partial \phi} \Bigl] 
\nonumber \\
& & -8\pi G_4 \tau-{\kappa_5}^4 {\pi^\mu}_\mu \label{trace}.
\end{eqnarray}
%
Hereafter we set ${\kappa_5}^2=1$ for the brevity. 

\section{View from the Holography}\label{III}

\subsection{Set-up}

To derive the effective equation on the brane from the holographic 
point of view, we follow the Giddings, Katz and Randall's argument \cite{Lisa},
where they started with the following observation in the path integral:
%
\begin{eqnarray}
Z & = & \int {\cal D}g e^{i[S_5(g)+\frac{1}{2}S_{\rm brane}(q)]} \nonumber \\
  & = &  \int {\cal D}q e^{\frac{i}{2}S_{\rm brane}}
\int_{g|_b=q} {\cal D}g e^{iS_5} \nonumber \\
  & = &  \int {\cal D}q e^{\frac{i}{2}S_{\rm brane}} 
e^{iS_{\rm ct}}\Bigl\langle 
e^{i \int d^4x q_{\mu\nu}T^{\mu\nu}} \Bigr\rangle_{\rm CFT}.
\label{adscft}
\end{eqnarray}
%
In the above a coupling of the bulk scalar to 
a boundary operator is tacitly included, this is, $g$ expresses 
the both of the metric and scalar for the brevity.   
$S_5$ is the gravitational action with the boundary term, 
$S_{\rm brane}$ is the brane action including the brane tension and 
the matter fields on the brane. When we move from the second to the third lines, 
we have used the AdS/CFT correspondence\cite{AdSCFT};
Roughly speaking, the classical gravity in 
the bulk is dual with CFT living on the boundary. 
Then, we must introduce the counter-term $S_{ct}$ to make the 
action finite. The counter-term is assumed to have the 
local form in terms of the 4-dimensional quantities, which  
can be easily determined by using the Hamilton-Jacobi equation \cite{HJ}
(See also \cite{Jiro})\footnote{The Hamilton-Jacobi equation 
was applied into cosmology to discuss the inhomogeneities in the 
long wave limit\cite{GE}. 
The long wave approximation seems to correspond to the 
low energy limit.}:
%
\begin{eqnarray}
{\cal H} & = & \frac{2}{{\sqrt {-q}}}
\frac{\delta S}{\delta q_{\alpha\beta}} \frac{\delta S}{\delta q_{\mu\nu}}
\Bigl(q_{\alpha\mu}q_{\beta\nu}-\frac{1}{3}q_{\alpha\beta}q_{\mu\nu} \Bigr)
\nonumber \\
& & 
+\frac{1}{2}\frac{1}{{\sqrt {-q}}} \Bigl( \frac{\delta S}{\delta \phi}\Bigr)^2
-\frac{1}{2}{\sqrt {-q}}(D \phi)^2+\frac{1}{2}{\sqrt {-q}}{}^{(4)}R 
\nonumber \\
& & -{\sqrt {-q}}V(\phi)=0.
\end{eqnarray}
%
This will be solved by expanding on shell
action $S$ with respect to the order of the 
derivatives, that is, we perform the derivative expansion.

\subsection{An example}

In this subsection, we will show an explicit relation between 
the ``electric'' part of the 5-dimensional Weyl tensor and the 
energy-momentum tensor of  CFT living on the brane to the 
2nd order of the derivative expansion. 
To the 4th order, we will 
obtain the 
relation between ${\pi^\mu}_\mu$ and the trace-anomaly of  CFT.   

First of all, we remind that the generalized AdS/CFT argument 
depends on the scheme in the general bulk potential except for 
the trivial case. Apart from the Hamilton-Jacobi formalism adopted 
in this report, there is another way where we take the metric expansion  
near the brane (i.e., boundary) \cite{Ogushi}. 
To obtain the universal result, our attention 
will be focused on the constant potential cases. 

We here consider the system which can be controlled by 
the counter-term
%
\begin{eqnarray}
S_{\rm ct}& = & \frac{3}{\ell}\int d^4x{\sqrt {-q}}
+\frac{\ell}{4}\int d^4 x{\sqrt {-q}}
({}^{(4)}R-D_\mu \phi D^\mu \phi) \nonumber \\
& & +S^{(4)}+\cdots,
\label{ct1}
\end{eqnarray}
%
where $S^{(4)}$ is the counter term including the 4th derivatives like 
${}^{(4)}R^2$ and ${}^{(4)}R(D\phi)^2$. As seen soon, the system must have the 
trivial potential for the bulk scalar field. 

{}From the Hamilton-Jacobi equation 
of the 0th order, we obtain 
%
\begin{eqnarray}
V(\phi)=-\frac{6}{\ell^2}={\rm constant}.\label{poten}
\end{eqnarray}
%
Using the above counter-term in Eq.~(\ref{adscft}), 
we obtain the effective gravitational equation on the brane:
%
\begin{eqnarray}
{}^{(4)}G_{\mu\nu} & = & 
8\pi G_4 \tau_{\mu\nu}+D_\mu \phi D_\nu \phi -\frac{1}{2}q_{\mu\nu}
(D \phi)^2 \nonumber \\
& & +\frac{4}{\ell}\frac{1}{{\sqrt {-q}}}
\frac{\delta S^{(4)}}{\delta q_{\alpha\beta}}q_{\mu\alpha}q_{\nu\beta}
+\frac{4}{\ell}\langle T_{\mu\nu} \rangle_{\rm CFT}.\label{holo1}
\end{eqnarray}
%
This equation should be compared with Eq.~(\ref{basic1}). 
In the above we set the net cosmological constant on the
brane to be zero, which requires
%
\begin{eqnarray}
\lambda=\frac{6}{\ell}.
\end{eqnarray}
%
Comparing 
 Eq.~(\ref{basic1}) with Eq.~(\ref{holo1}), we obtain the 
relation between $E_{\mu\nu}$ and the energy-momentum tensor of 
CFT in the 2nd derivative order or $O(T_{\mu\nu})$:
%
\begin{eqnarray}
E_{\mu\nu}& = &-\frac{4}{\ell}\langle T_{\mu\nu}
-\frac14 q_{\mu\nu}T\rangle_{\rm CFT}  \nonumber
\\
& & -\frac{1}{3}\left[D_\mu \phi D_\nu \phi
-\frac{1}{4}q_{\mu\nu}(D \phi)^2 \right]. \label{Eads}
\end{eqnarray}
%
Although $\langle T_{\mu\nu} \rangle_{\rm CFT}$ is expected to be 
higher order than 
$(D\phi)^2$ terms, we keep it in the right-hand side in Eq.~(\ref{Eads}). 

The trace of the effective equation is 
%
\begin{eqnarray}
{}^{(4)}R=-8\pi G_4 \tau +(D\phi)^2-\frac{4}{\ell}\langle {T^\mu}_\mu 
\rangle_{\rm CFT} \label{trace2},
\end{eqnarray}
%
where we used the fact 
that the tensor produced from the 4th order local term $S^{(4)}$ 
is traceless:
%
\begin{eqnarray}
\frac{\delta S^{(4)}}{\delta q_{\mu\nu}} q_{\mu\nu}=0.
\end{eqnarray}
%
The trace-anomaly of CFT can also be evaluated by using 
the Hamilton-Jacobi equation \cite{HJ,Ano,Ogushi,Nojiri,Fukuma,Old}:
%
\begin{eqnarray}
\langle {T^\mu}_\mu \rangle_{\rm CFT} 
 & = & \frac{\ell^3}{16}\Bigl[{}^{(4)} R_{\mu\nu}{}^{(4)}R^{\mu\nu}
-\frac{1}{3}{}^{(4)}R^2 \nonumber \\
& & -2{}^{(4)}R^{\mu\nu}D_\mu \phi D_\nu \phi 
 +\frac{2}{3}D_\mu \phi D^\mu \phi {}^{(4)}R \nonumber \\
& & +\frac{2}{3}(D_\mu \phi D^\mu \phi)^2
\Bigr]. \nonumber \\ \label{coupling}
\end{eqnarray}
%

Let us compare Eq.~(\ref{trace}) with Eq.~(\ref{trace2}). To do so 
we set $\Lambda_4=0$ in Eq.~(\ref{trace}). Since $\lambda$ is 
constant, Eq.~(\ref{trace}) becomes 
%
\begin{eqnarray}
{}^{(4)}R  =  -8\pi G_4 \tau +(D\phi)^2 
 -\frac{1}{4}
\Bigl(\frac{\partial {\cal L}_m}{\partial \phi}  \Bigr)^2 
- {\pi^\mu}_\mu. \label{trace3}
\end{eqnarray}
%
Using the Einstein equation to the 2nd order, 
${}^{(4)}G_{\mu\nu} \simeq 8\pi G_4 \tau_{\mu\nu}+D_\mu \phi D_\nu \phi
-\frac{1}{2}g_{\mu\nu}(D\phi)^2$, 
together with Eqs.~(\ref{trace2}) and (\ref{trace3}) implies 
%
\begin{eqnarray}
\frac{\partial {\cal L}_m}{\partial \phi}=0.
\end{eqnarray}
%
This means that the coupling of the 
scalar field to the matter fields on the brane is not
admitted. It is reminded that the holographic CFT does 
couples to the scalar field. 

As a summary, we have shown
%
\begin{eqnarray}
{\pi^\mu}_\mu=\frac{4}{\ell}\langle {T^\mu}_\mu \rangle_{\rm CFT}.
\end{eqnarray}
%
It is already well-known that 
the term, ${\pi^\mu}_\mu$, implies the modification of the 
FRW models\cite{cosmos}:
%
\begin{eqnarray}
\Bigl( \frac{\dot a}{a}\Bigr)=\frac{8\pi G_4}{3}\rho+\frac{1}{36}
\rho^2.
\end{eqnarray}
%
Hence we can conclude that 
$\rho^2$-term comes from the trace-anomaly from the holographic 
point of view. 

\section{Summary}\label{IV}

We considered the generic features of the brane-world 
from the holographic point of view. We showed that the explicit relation 
between the ``electric'' part of the Weyl tensor and the energy-momentum 
tensor of  CFT living on the brane. This gives us the reformulation 
of previous works of Refs. \cite{Lisa,AdS1} in the system with the bulk scalar 
field. In addition, the $\rho^2$-term comes from the trace anomaly of CFT. 

We stress that the brane-world models  
with a correct AdS/CFT interpretation 
will belong to a rather limited class in general.
As an example, 
we considered the bulk massless scalar field and 
showed that the coupling of the scalar field with the matters on the 
brane cannot be admitted. This is also 
one of our main results. On the other hand, it is
reminded that holographic CFT is coupled to the bulk
scalar on the branes(See Eq.~(\ref{coupling})).

\section*{Acknowledgements}

We would like to thank T. Kawano, S. Nojiri, H. Ochiai and Y. Shimizu for 
fruitful discussions. 

\appendix

\section{Derivation of the counter terms}

In general the 0th order of the counter terms can be 
written as 
%
\begin{eqnarray}
S^{(0)}=3\int d^4 x {\sqrt -q}H(\phi).\label{A1}
\end{eqnarray}
%
In this order the Hamilton-Jacobi equation is 
%
\begin{eqnarray}
& & \frac{2}{{\sqrt {-q}}}
\frac{\delta S^{(0)}}{\delta q_{\alpha\beta}} 
\frac{\delta S^{(0)}}{\delta q_{\mu\nu}}
\Bigl(q_{\alpha\mu}q_{\beta\nu}-\frac{1}{3}q_{\alpha\beta}q_{\mu\nu} \Bigr)
\nonumber \\
& & \;\;\;\;\;\;\;
+\frac{1}{2}\frac{1}{{\sqrt {-q}}} 
\Bigl( \frac{\delta S^{(0)}}{\delta \phi}\Bigr)^2
-{\sqrt {-q}}V(\phi)=0.\label{A2}
\end{eqnarray}
%
Substituting Eq.~(\ref{A1}) into Eq.~(\ref{A2}), we obtain the relation 
between the potential for the bulk scalar and $H(\phi)$: 
%
\begin{eqnarray}
V(\phi)=\frac{9}{2}\Bigl[\Bigl( \frac{d H}{d\phi}\Bigr)^2
-\frac{4}{3}H^2(\phi)  \Bigr].
\end{eqnarray}
%
Since $H(\phi)=1/\ell$ in the text, we obtain Eq. (\ref{poten}). 
In the 2nd order the Hamilton-Jacobi equation is 
%
\begin{eqnarray}
& & \frac{4}{{\sqrt {-q}}}
\frac{\delta S^{(0)}}{\delta q_{\alpha\beta}} 
\frac{\delta S^{(2)}}{\delta q_{\mu\nu}}
\Bigl(q_{\alpha\mu}q_{\beta\nu}
-\frac{1}{3}q_{\alpha\beta}q_{\mu\nu} \Bigr)
+\frac{1}{{\sqrt {-q}}} 
\frac{\delta S^{(0)}}{\delta \phi}\frac{\delta S^{(2)}}{\delta \phi}
\nonumber \\
& & \;\;\;\;\;\;\;
-\frac{1}{2}{\sqrt {-q}}(D \phi)^2+\frac{1}{2}{\sqrt {-q}}{}^{(4)}R=0.
\end{eqnarray}
%
Substituting Eq. (\ref{A1}) into the above, we obtain Eq. (\ref{ct1}) in 
the text.


\end{document}